\documentclass[11pt,preprint2]{aastex}

\slugcomment{To appear in APJS}

\shorttitle{The Robustness of LSFS}
\shortauthors{Winkel \& Kerp}
\usepackage{amsmath,amsfonts}

\begin{document}


\title{The Robustness of Least-Squares Frequency Switching (LSFS)}


\author{B. Winkel$^\mathrm{1,2,}$\altaffilmark{a} and J. Kerp$^\mathrm{2}$}
\affil{$^\mathrm{1}$Max-Planck-Institut f\"{u}r Radioastronomie, Auf dem H\"{u}gel 69, 53121 Bonn, Germany}
\affil{$^\mathrm{2}$Argelander-Institute for Astronomy (AIfA), University of Bonn,\\  Auf dem H\"{u}gel 71, D-53121 Bonn, Germany}
\email{bwinkel@astro.uni-bonn.de}
\email{jkerp@astro.uni-bonn.de}


\altaffiltext{a}{Member of the International Max Planck Research School (IMPRS) for Radio and Infrared Astronomy at the Universities of Bonn and Cologne}


\begin{abstract}
Least-squares frequency switching (LSFS) is a new method to reconstruct signal and gain function (known as bandpass or baseline) from spectral line observations using the frequency switching method. LSFS utilizes not only two but a set of three or more local oscillator (LO) frequencies. The reconstruction is based on a least squares fitting scheme. Here we present a detailed investigation on the stability of the LSFS method in a statistical sense and test the robustness against radio frequency interference (RFI), receiver gain instabilities and continuum sources. It turns out, that the LSFS method is indeed a very powerful method and is robust against most of these problems. Nevertheless, LSFS fails in presence of RFI signals or strong line emission. We present solutions to overcome these limitations using a flagging mechanism or remapping of  measured signals, respectively. 
\end{abstract}





\section{Introduction}
Radio astronomers were quite fast from the beginning confronted with problems as unknown intermediate frequency (IF) gain functions and instabilities, both in frequency and time, which makes spectral analysis of faint or very broad emission line sources very complicated. Up to now mainly two different schemes are used to overcome these problems: position switching and frequency switching. Both methods reduce the receiver gain instabilities by measuring a reference spectrum, either off-source (position switching) or with a detuned local oscillator (LO) frequency to move the line of interest outside the observed spectral band (frequency switching). There are a lot of drawbacks applying both methods. Using position switching it is necessary to avoid different continuum levels (or very broad line emission) towards the on and off position which of course limits the usage of position switching in Milky Way observations, having emission from all directions. This is especially valid for the important spectral lines of neutral atomic Hydrogen or CO, due to their high area filling factor. Furthermore, there is loss of observing time either when redirecting the telescope between on and off positions or during the retuning of the LO frequency, respectively. Frequency switching  suffers also from the gain curve changes while shifting the spectral range. The most problematic aspect of both methods is the loss of half of the observing time. In-band frequency switching (both LO phases provide the line of interest lying in the observed bandwidth) avoids this loss. Unfortunately, one loses half of the available bandwidth (which is equivalent to the loss of velocity coverage or number of spectral channels), because a proper separation of both signals is needed. Observing single objects this is usually not a major problem, but when performing, for example, a blind survey one would strongly suffer from such a restriction. We also point out, that many important questions of modern cosmology rely heavily on highest-sensitivity radio measurements of the high-redshift universe. Modern spectrometers based on field programmable gate arrays (FPGA) technology \citep[e.g.][]{benz05,stanko05} are best suited for that purpose but meaningless without sophisticated methods for bandpass calibration.

\cite{heiles07} presented a new method called Least-Squares Frequency Switching (LSFS). LSFS is able to deal with all problems discussed above, making it the best choice for future spectral line observations with radio telescopes especially well suited for \ion{H}{1}. However, it requires minor hardware changes at the telescope in order to provide not only two, but a set of three or more LO frequencies within one switching cycle. This is not a substantial problem, and there already is a working system using LSFS at the Arecibo telescope \citep{heiles07,stani06}. 

Of major interest in radio spectral observations --- especially in high-redshift \ion{H}{1} astronomy --- is the ability of a ``bandpass removal'' tool to provide high-quality results in a statistical sense. Most observations today need to integrate at least a couple of minutes to reach the desired sensitivity limit. While in theory the noise level scales as $1/\sqrt{t\Delta\nu }$ according to the radiometer equation, with $\Delta\nu$ being the bandwidth and $t$ the integration time, this is not necessarily true for a real receiving system. Modern backends have Allen times, $t_0$, (for $t\leq t_0$ the radiometer equation holds) of hundreds of seconds \citep{stanko05}. \cite{winkel07} show that in presence of radio frequency interferences (RFI) this can be limited to less than few tens of seconds. We analyze, if the LSFS method has an impact on the RMS level (the sensitivity) achievable. It turns out, that in statistical sense the LSFS performs very well, yielding only a slightly decrease of sensitivity. We also test the robustness of the LSFS versus several typical problems at radio telescope sites. These are for example RFI events, possibly bandpass instabilities in time and frequency, as well as (strong) continuum sources. LSFS works well under most tested circumstances except for RFI and strong emission lines. However, small changes to the original LSFS method already provide meaningful workaround mechanisms. These are discussed in subsequent sections.

Section~\ref{secmethodlsfs} contains a brief summary about the LSFS method based on \cite{heiles07}. The robustness of LSFS is analyzed in Section~\ref{seclsfsrobustness}. In Section~\ref{seccomputingeff} we finally calculate the computing time needed in order to perform the various calculations related to the LSFS method. We propose to use a modified method for the computation of the SVD, based on an algorithm available for sparse matrices. This improves computation speed for the (one-time per LO setup) calculation of the correlation matrix by an order of magnitude. Section~\ref{secsummary} contains our summary.

\section{Least-squares frequency switching (LSFS)}\label{secmethodlsfs}
Most spectroscopic observations in radio astronomy use the heterodyne principle where the radio frequency (RF) signal is multiplied with a monochromatic signal of a LO. An appropriate low-pass filter applied after this operation provides the desired IF signal at much lower carrier frequencies. The whole system can be described by
\begin{equation}
\begin{split}
 P_\mathrm{IF}&(f_\mathrm{IF})= G_\mathrm{IF}(f_\mathrm{IF})G_\mathrm{RF}(f_\mathrm{RF})\times \\
 &
\left [ T_A(f_\mathrm{RF})+T_A+T_R(f_\mathrm{RF})+T_R\right]\label{eqbasicequation}
\end{split}
\end{equation}
with $P_\mathrm{IF}$ being the power in the IF chain. $G_\mathrm{IF}$ and $G_\mathrm{RF}$ are the gain functions at IF and RF stage, respectively. The gain acts on the signals which enter the feed --- the astronomical signal of interest plus the contribution from the sky which we denote as  $T_A$ --- as well as on the noise of the receiver, $T_R$. \cite{heiles07} separates $T_A$ and $T_R$ into a frequency dependent and independent (continuum) part. Note that the gain is not simply a scalar but has a spectrum due to the filter curves. 

To recover from the measured signal, $P_\mathrm{IF}$, the signal of interest, $T_A$, one needs to know the gain spectrum $G_\mathrm{IF}$ ($G_\mathrm{RF}$ can be treated as constant with frequency on modern receivers). Traditionally, this is achieved by measuring a reference spectrum without any spectral features either by position or frequency switching. We refer to \cite{heiles07} for a review of position and frequency switching.

Equation~(\ref{eqbasicequation}) reads in a simplified form as
\begin{equation}
P_\mathrm{IF}(f_\mathrm{IF})=G_\mathrm{IF}(f_\mathrm{IF})S_\mathrm{RF}(f_\mathrm{RF})
\end{equation}
where we combined all signals entering the mixer to $S_\mathrm{RF}(f_\mathrm{RF})$.  In contrast to \cite{heiles07} we dropped the assumption that the input signals are a superposition of frequency dependent and independent parts. Using modern broadband spectrometer backends we likely can not treat the continuum signal as constant over the entire observed bandwidth. Moreover we might also be interested  in the continuum emission itself. Broadband spectrometers could also be used to generate continuum maps. Furthermore, from the mathematical point of view it is not necessary to perform the separation --- all subsequent equations do not rely on it.

As before, we are interested in obtaining $G_\mathrm{IF}(f_\mathrm{IF})$. Introducing not only 2 but a set of $N$ different LO frequencies \citep[for a detailed analysis of how to choose appropriate LO frequencies, see][]{heiles07}, we end up having $N\cdot I$ equations
\begin{equation}
 P_{i,\Delta i_n}=G_iS_{i+\Delta i_n}.
\end{equation}
In this representation we now use integer indices representing the spectral channels of the backend. Then $i$ is the $i^\mathrm{th}$ out of $I$ channels, $\Delta i_n$ is the frequency shift of LO~$n$ versus LO~$0$ (the unshifted LO) given in channels. By using different LO frequencies we of course observe somewhat different spectral portions of the input spectrum $S$. Without loss of generality we normalize the input signal to have a mean value of unity, $S_{i+\Delta i_n}=1+s_{i+\Delta i_n}$, leading to 
\begin{equation}
 P_{i,\Delta i_n}=G_i+G_is_{i+\Delta i_n}\label{eqnonlinear}
\end{equation}
which can be solved using nonlinear least-squares techniques. However, \cite{heiles07} converted the equation to an iterative linear least-squares problem by solving for the difference of guessed values of $G_i^\mathrm{g}$ and $s_{i+\Delta i_n}^\mathrm{g}$ from their true values. From these guessed values one can of course compute the associated output $P_{i,\Delta i_n}^\mathrm{g}$ power for each spectral channel and LO setting. After some simplifications \citep[dropping higher order terms; see][]{heiles07} equation~(\ref{eqnonlinear}) transforms  into
\begin{equation}
\frac{\delta P_{i,\Delta i_n}}{G_i^\mathrm{g}}=\frac{\delta G_i}{G_i^\mathrm{g}} 
 +\delta s_{i+\Delta i_n}\label{eqlinear}.
\end{equation}
The $\delta$-terms denote the difference between the true and the guessed value of the corresponding quantity. A further constraint is needed in order to keep the mean RF power approximately constant, namely $\sum_{i,n}\delta s_{i+\Delta i_n}=0 $.

For convenience we use matrix notation for equation~(\ref{eqlinear})
\begin{eqnarray}
&  \mbox{\boldmath $p$}= \mathbf{X} \mbox{\boldmath $a$}\\
& \mbox{\boldmath $p$}^\mathrm{T}\equiv\left( \mbox{\boldmath $p$}_{i,0}^\mathrm{T},\ldots, \mbox{\boldmath $p$}_{i,N-1}^\mathrm{T} \right)\label{eqlinearmatrix} \\ 
&\mbox{\boldmath $a$}^\mathrm{T}\equiv\left( g_0, \ldots g_{I-1}, \delta s_0,\ldots,\delta s_{I-1+\Delta i_{N-1}} \right)\\
& p_{i,n}\equiv \displaystyle\frac{\delta P_{i,\Delta i_n}}{G_i^\mathrm{g}},\qquad g_i\equiv\frac{\delta G_i}{G_i^\mathrm{g}}.
\end{eqnarray} 
Least-squares fitting is achieved by computing
\begin{equation}
 \mbox{\boldmath $a$}=\left( \mbox{\boldmath $\alpha$}\, \mathbf{X}^\mathrm{T} \right)\mbox{\boldmath $p$}, \qquad \mbox{\boldmath $\alpha$}\equiv\left( \mathbf{X}^\mathrm{T}\mathbf{X}   \right)^{-1}
\end{equation}
with the covariance matrix $\mbox{\boldmath $\alpha$}$. Computing $\mbox{\boldmath $\alpha$}$ requires matrix inversion which in general does not exist necessarily. To deal with degeneracies, \cite{heiles07} proposes the Singular-Value Decomposition (SVD) of matrix $\mathbf{X}$
\begin{equation}
 \mathbf{X}=\mathbf{U}[\mathbf{W}]\mathbf{V}^\mathrm{T}
\end{equation}
with the diagonal matrix $\mathbf{W}$ containing the so-called singular values $w_i$. In the case of degeneracies one or more of the $w_i$ are close to zero, leading to infinite (or huge) numbers when inverting. It turns out that
\begin{equation}
 \left( \mbox{\boldmath $\alpha$} \mathbf{X}^\mathrm{T}\right)=\mathbf{V}\left[ \frac{1}{\mathbf{W}} \right] \mathbf{U}^\mathrm{T}.
\end{equation}
The critical singular values can be treated separately (e.g. setting the inverse values to zero). By computing the SVD of the matrix $\mathbf{X}$ one can directly solve equation~(\ref{eqlinearmatrix}) without encountering any problems caused by degeneracies. The computation of the SVD of a matrix, e.g. for $N=8,~I=1024$, is possible on a modern PC but is not finished within fractions of a second; see Sect.\,\ref{seccomputingeff} for details. Nevertheless, the SVD calculation fortunately needs to be done only once per LO setup, as the matrix itself is independent from the measurements. 

The assumed normalization of the signal seems to be a somewhat arbitrary assumption. But if we assume the bandpass (gain) curve to be normalized, we can in practice easily attribute associated gain factors to the signal which nevertheless has to be calibrated in terms of intensity. This way we can uniquely reconstruct the overall power of the input signal (in arbitrary units) by computing the mean, $m$, of the measured signal, normalizing (dividing by $m$), calculating the LSFS which gives a signal of mean value of unity and finally multiplying the reconstructed signal with $m$. It is clear that this scheme will only work if the gain curve remains constant. This can be expected at least for the duration of the observing session so that the computed gain factor remains constant for a single observation. 

The LSFS method also allows reconstructing the continuum part of the input signal. While for position and frequency switching the separation of RF-dependent and -independent parts was necessary, the LSFS algorithm allows the reconstruction of the complete mixture of signals which are put into the mixer. Of course this implies that there is possibly the need for further disentangling these signals into line and continuum components (from astronomical sources, ground, and receiver noise) which may even have different spectral indices. 

The equations hold for small values of $s_{i+\Delta i_n}$ as they were computed only to first order approximation. In most cases this is easily fulfilled in radio astronomy as the observed lines are much weaker than the typical intensity of the unavoidable continuum level produced by the atmosphere, ground, and receiver noise, which sum is known as the system temperature $T_\mathrm{sys}$. However, there is one case known where we indeed have a signal much brighter than the continuum level: the \ion{H}{1} emission of the Milky Way which can reach intensities of a few 100\,K while the system temperature  for a typical telescope is $20-40$\,K. We will address this issue in the following analysis.

\section{Robustness of the LSFS}\label{seclsfsrobustness}
\subsection{Setup}
We implemented the LSFS algorithm within the programming language C and performed various tests to investigate statistical stability, response to possible variations of bandpass shapes, impact of RFI signals, and its ability to deal with (strong) continuum sources. 

For our testing purposes we generated spectra ($1024$ spectral channels) by simulating several Gaussian-shaped (faint) emission lines of different intensities and widths on top of a constant signal (which shall resemble those continuum signals with spectral index of zero). After adding Gaussian noise, these emission lines are partly not anymore visible; compare for example the signal spectrum of Fig.\,\ref{figsinglespec} and Fig.\,\ref{figintegratedspec}. This ``true'' signal is then multiplied with a  gain function 
\begin{eqnarray}
&G_\mathrm{IF}(f_\mathrm{IF})=G_\mathrm{IF}^\mathrm{filt}\cdot G_\mathrm{IF}^\mathrm{wave} \cdot G_\mathrm{IF}^\mathrm{poly}\\
&\rule{-3ex}{0ex}G_\mathrm{IF}^\mathrm{filt}=\frac{1}{2}\left[\tanh\left(5f+5\right)-\tanh\left(5f-5\right)\right]\\
&G_\mathrm{IF}^\mathrm{wave}=1+0.1\cos\left(F\pi f\right)\\
&G_\mathrm{IF}^\mathrm{poly}=1+Af+0.5f^2
\end{eqnarray}
which we adopted from \cite{heiles07} to allow for a better comparison of our results to that work. $f$ is the frequency which we transformed to the spectral ($i$) domain using $f=2.1(i-512)/1024$. In contrast to \cite{heiles07} we chose a spectral portion where the gain curve does not gets too close to zero. This would break the assumption that $G_i$ is of order unity and distorts the normalization scheme which we presented in the previous section. \cite{heiles07} did not encountered that problem because he neglects continuum emission. Note, that this is no drawback, as one can easily choose in practice those portions of the spectra which fulfill $G_i\simeq1$. We varied the two parameters around $A=0.1$ and $F=4$ to change the bandpass in amplitude and shape for some of our tests. A small variation of $F$ around $4$ already has a dramatic impact on overall shape of the gain curve.

\subsection{Statistical stability}

\begin{figure}[!t]
\plotone{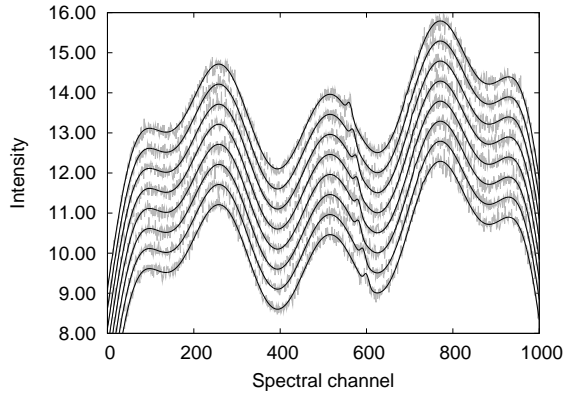}
\caption{Raw input spectra as would be measured by the receiving system using the MR8 scheme. For better visualization the spectra (grey solid lines) were stacked and a noise-free analogon (black solid lines) was overplotted. Each spectrum is the multiplication of the ``true'' input signal and the IF gain function. Due to the different LO frequencies within a LO cycle the signals of interest are folded to different spectral channels.}
\label{figrawinput}
\end{figure}

\begin{figure}[!t]
\plotone{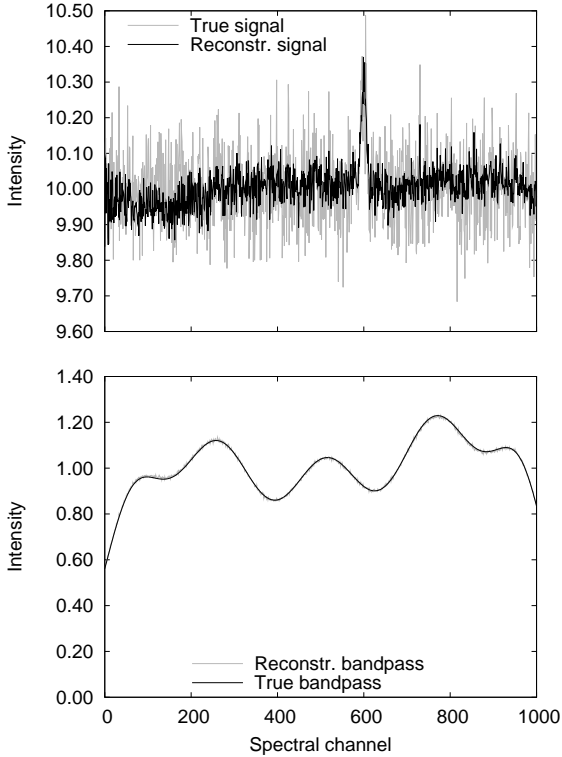}
\caption{Reconstructed and original signal (top) and bandpass shape (bottom) of a single spectrum from our simulations. The signal is a superposition of noise and three unrelated line signals --- two of them are well below the noise level but are visible after integration of several spectra; see Fig.\,\ref{figintegratedspec} (top). Note that the noise level of the reconstructed signal is about a factor $\sqrt{8}$ smaller because 8 spectra (the different switching phases) result in a single reconstructed signal spectrum. }
\label{figsinglespec}
\end{figure}

We started by examining the statistical stability of the method. For this purpose we generated spectra for a set of 8 LO frequencies using the MR8 scheme \citep{heiles07}; see Fig.\,\ref{figrawinput}. As shown in Fig.\,\ref{figsinglespec}, the solution for a single set of spectra does not necessarily provides the ``true'' signal and bandpass shape, but there are small systematic effects. We attribute these partly to the influence of single (strong) noise peaks to the overall solution. Remember also, that the set of equations which we use to solve the decoupling of signal and gain is in linear-order approximation. It is important to note, that the iteration was in any case not interrupted until the solution had converged. The question is whether these systematics cancel out after integration of several spectra or whether they remain. In order to have a measure of the ``goodness'' of the solution we adopt from \cite{heiles07} two quality indicators --- the  \textit{RMS level of the reconstructed signal} (denoted as RMS) and the \textit{RMS of the residual gain curve} (denoted as $\sigma$). The latter quantity uses the residual which is the difference between the true gain curve (noise-free) and the reconstructed gain. The first quality indicator, the RMS, is calculated making use of all spectral channels except those containing the signals of interest. For the purpose of comparison with the noise level of the signals we also rescale $\sigma$ with the gain factor, which was used to scale the signals. It is also obvious, that the reconstructed signals ideally should have a factor of $\sqrt{8}$ lower noise compared to the originally generated signals. This is because each reconstructed signal was calculated using eight ``observed spectra'' (one LO cycle). Furthermore, we analyze the behavior of the indicators as a function of integration `time'. This is done by successively summing up adjacent spectra. In each step this reduces the number of spectra by a factor two. Therefore, we start with a total number of generated 1024 true spectra (or $8\cdot1024$ measured spectra, respectively) which is a power of two. The LSFS was calculated for each of the 1024 spectra, then the summation of the reconstructed signal and gain curve was performed stepwise. Theoretically, the functional dependence of RMS vs. integration time is given by the radiometer equation
\begin{equation}
P(t)\sim \frac{1}{\sqrt{\Delta\nu\cdot t}}\sim t^{-0.5}.
\label{eqradiometer}
\end{equation}

\begin{figure}[!t]
\plotone{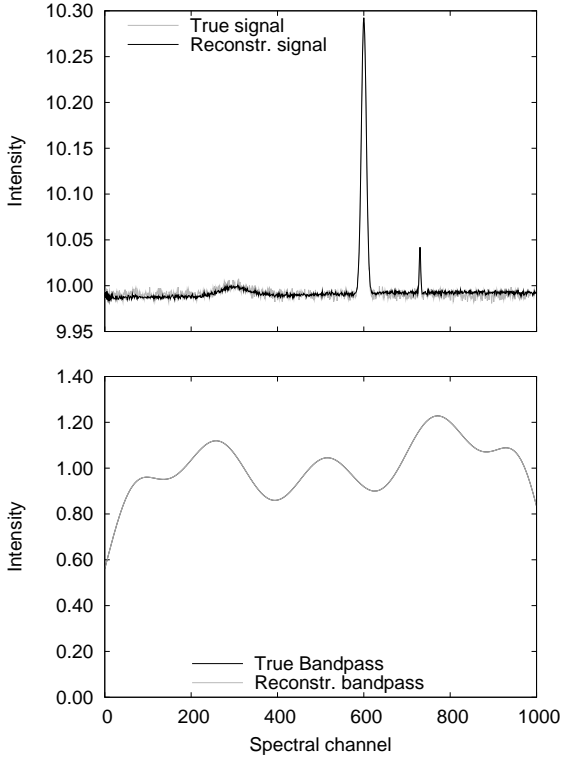}
\caption{After integration of 1024 subsequent spectra the faint signals previously hidden by noise show up. Both signal (top) and bandpass shape (bottom) are well recovered, as the functional behavior of the quality indicators in Fig.\,\ref{figsensitivity} reveals.  }
\label{figintegratedspec}
\end{figure}

\begin{figure}[!t]
\plotone{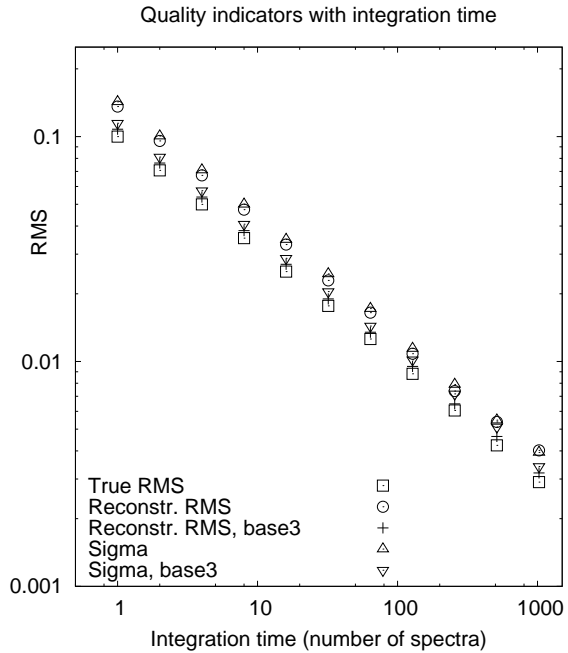}
\caption{Functional dependence of the different quality indicators vs. integration time (scans). The boxes mark the noise level (RMS) of the true signal, the circles represent the noise of the reconstructed signal. The triangles (up) mark the RMS values of the gain residual, $\sigma$, calculated using the difference of the true and reconstructed gain curves. After subtracting a third-order polynomial both the RMS (crosses) and $\sigma$ (triangles, down) quantities are closer to the theoretical value.}
\label{figsensitivity}
\end{figure}

Fig.\,\ref{figintegratedspec} clearly shows, that despite the fact that individual spectra were not perfectly handled the integrated signal as well as the bandpass visually match well. However, Fig.\,\ref{figsensitivity} reveals increased RMS values of about $30\%$ of the reconstructed signal and about $35\%$ higher noise for $\sigma$. Calculating the RMS and $\sigma$ values with respect to a 3rd-order polynomial (fitted after integration) results in significantly lower noise values which are only slightly increased  compared to the theoretical expectation value by $8\%$ (RMS) and $10\%$ ($\sigma$), respectively. Obviously the residual systematics can be described by a low-order polynomial. We point out, that the RMS behavior of the signal is of much greater interest from the observers point of view. If the gain curve is sufficiently stable with time, one can also compute the systems gain dependence to high precision by using thousands of spectra. The results show that one is effectively not losing sensitivity using this method as this was the case in earlier attempts \citep[e.g.][]{liszt97} with increased noise levels of about 100\%.

\subsection{LSFS and strong line emission}

\begin{figure}[!t]
\plotone{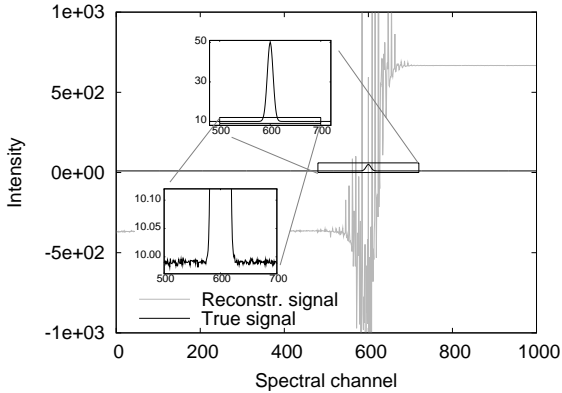}
\caption{In the presence of a strong emission line the LSFS method fails entirely. After normalization, all spectral features must be close to unity, otherwise the linear-order approximation is no longer valid. A solution to the problem is remapping of the input signal; see Fig.\,\ref{figstronglineremapping} and Fig.\,\ref{figstronglineremappingquality}.}
\label{figstrongline}
\end{figure}

In the previous section we addressed the possible problem of strong emission lines (as would be the case in galactic \ion{H}{1} observations) which could violate the assumption of small variations of the (normalized) signal around unity. Here, we use a strong emission line to test its influence on the LSFS. To make it short --- the LSFS fails completely; see Fig.\,\ref{figstrongline}. The intensity of the strong line signal at spectral channel 600 is about 5 times higher than the baseline level (system temperature). This causes heavy distortions during the reconstruction process (comparing the relative amplitudes of the input and reconstructed signal).

\begin{figure}[!t]
\plotone{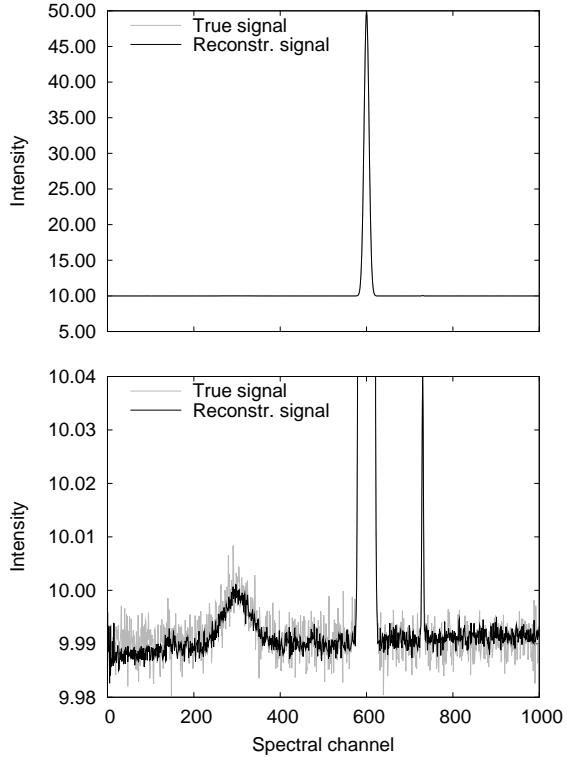}
\caption{Strong emission lines (top) can be handled by remapping the measured signal by a nonlinear function, e.g. $P\rightarrow\sqrt[4]{P}$. This ensures the LSFS method to be in the linear regime. The bottom panel shows a zoom-in for better visualization. The quality indicators are shown in Fig.\,\ref{figstronglineremappingquality}.}
\label{figstronglineremapping}
\end{figure}

\begin{figure}[!t]
\plotone{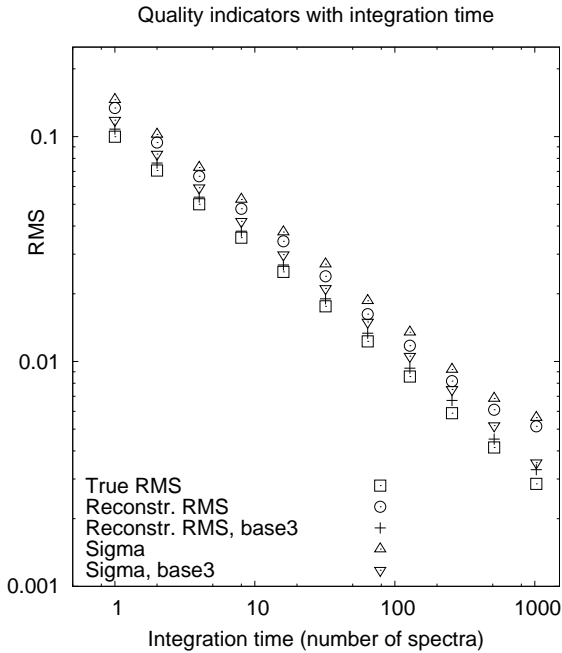}
\caption{Strong emission lines (top) can be handled by remapping the measured signal by a nonlinear function, e.g. $P\rightarrow\sqrt[4]{P}$. This ensures the LSFS method to be in the linear regime. The quality indicators (see Fig.~\ref{figsensitivity} for the explanation of the symbols) show that the remapping works correctly in a statistical sense. There is no significant increase of the RMS or $\sigma$ values compared to the undisturbed case; see Fig.\,\ref{figsensitivity}.}
\label{figstronglineremappingquality}
\end{figure}

We identified a workaround to the problem: by remapping the observed (normalized) signal, $P$, in terms of a nonlinear function one can treat strong signals into the realm of small variations around unity. In our example we tried the mapping function $P\rightarrow\sqrt[x]{P}$, with $x=4$. It is not clear, though, that the reconstructed signal and bandpass can be transformed back by simply using the inverse $P\rightarrow P^x$. But indeed it turns out, that this is possible; see Fig.\,\ref{figstronglineremapping} and Fig.\,\ref{figstronglineremappingquality}. 

\subsection{Bandpass instabilities}

\begin{figure}[!t]
\plotone{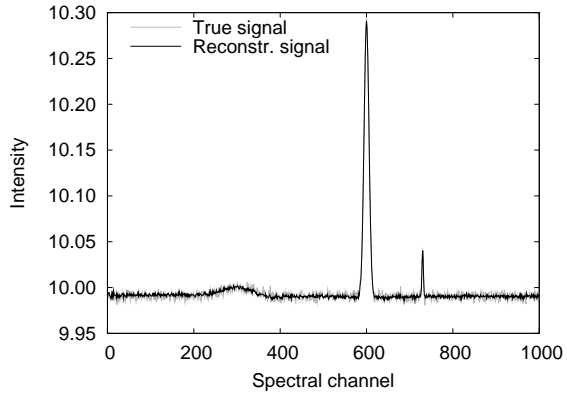}
\caption{A slow changing bandpass shape (constant gain curve during one LO cycle) has no measurable influence on the LSFS method. The signals can be treated as well recovered (see Fig.~\ref{figslowvaryingbpquality}).}
\label{figslowvaryingbp}
\end{figure}

\begin{figure}[!t]
\plotone{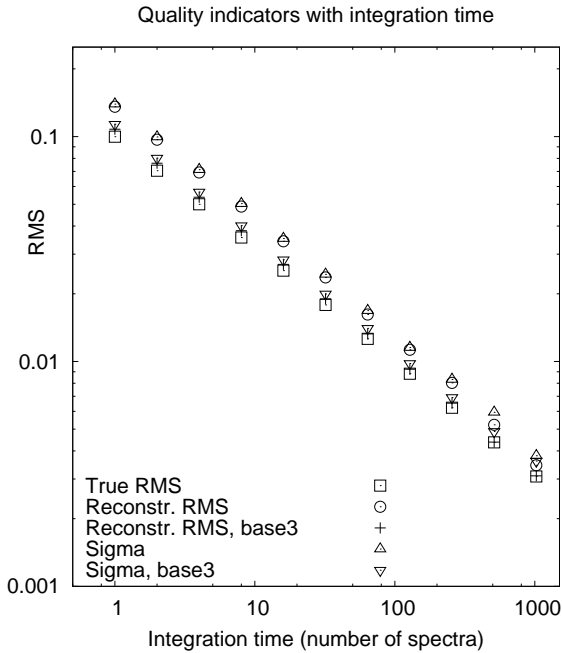}
\caption{A slow changing bandpass shape (constant gain curve during one LO cycle) has no measurable influence on the LSFS method. The quality indicators (see Fig.~\ref{figsensitivity} for the explanation of the symbols) show no significant increase of the RMS or $\sigma$ values compared to the undisturbed case; see Fig.\,\ref{figsensitivity}.}
\label{figslowvaryingbpquality}
\end{figure}

To further test the statistical stability, we now change the bandpass shape and amplitude with time. First, we only changed the shape slowly   but using the same shape for each bandpass within a single switching cycle (8 adjacent spectra have the same shape). This should resemble the situation at the telescope site as we can (hopefully) expect the bandpass shape to be independent of switching frequency, keeping in mind that the frequency shifts are very small compared to the total bandwidth. We could not find any significant difference to the undisturbed case; see Fig.\,\ref{figslowvaryingbp} and Fig.\,\ref{figslowvaryingbpquality}.

\begin{figure}[!t]
\plotone{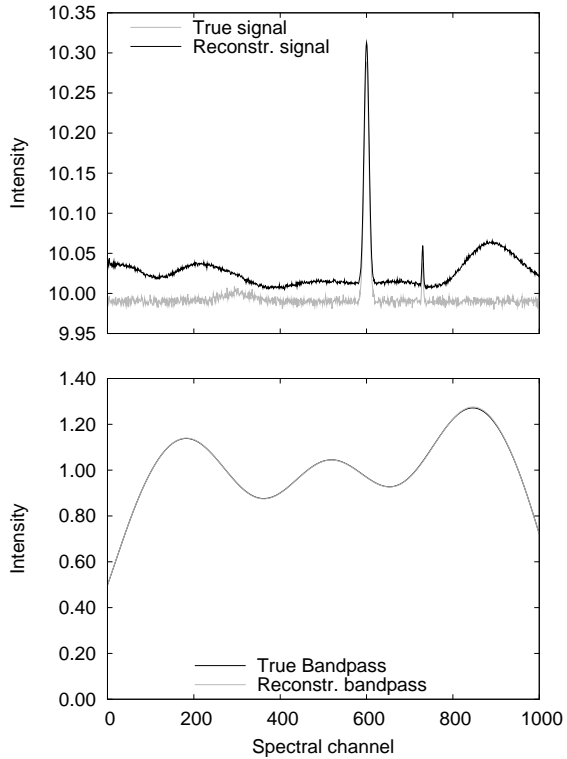}
\caption{A fast changing bandpass shape (see text) pushes the LSFS method to its limits. The bandpass and signal (top and bottom panel) were not well reconstructed. The difference between true and reconstructed gain curve as well as the baseline of the reconstructed signal can not be described by a low-order polynomial; see Fig.~\ref{figfastvaryingbpquality} for the quality indicators. Note, that due to the rescaling of the signal the uncertainties are much more visibly prominent in the signal domain than in the gain curves.}
\label{figfastvaryingbp}
\end{figure}

\begin{figure}[!t]
\plotone{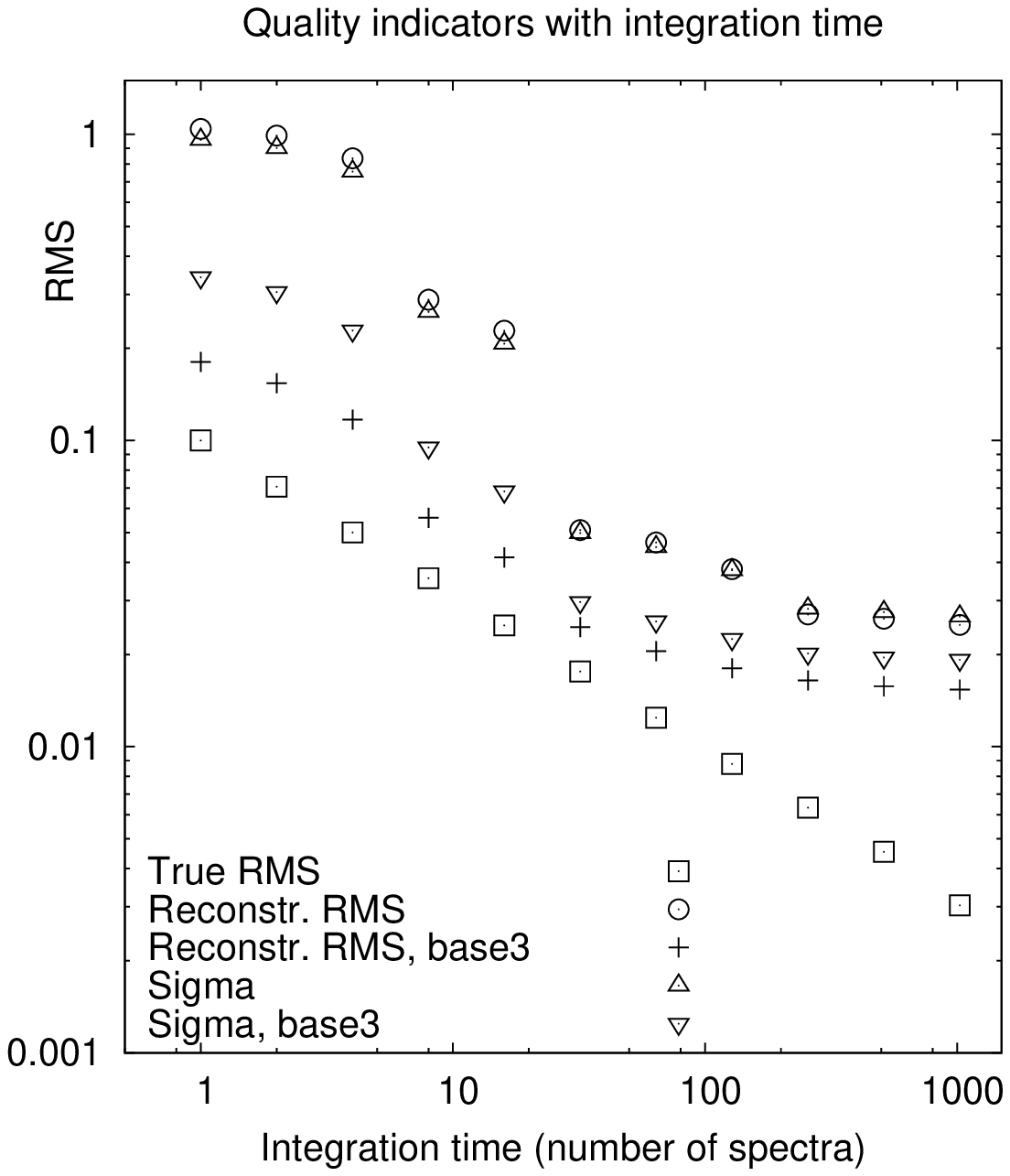}
\caption{Quality indicators (see Fig.~\ref{figsensitivity} for the explanation of the symbols) for a fast changing bandpass shape (see text). The difference between true and reconstructed gain curve as well as the baseline of the reconstructed signal can not be described by a low-order polynomial. Both the RMS and $\sigma$ values are much higher than in the undisturbed case and their functional behavior is far from linear.}
\label{figfastvaryingbpquality}
\end{figure}
As the slowly changing bandpass was no challenge for the LSFS algorithm, we also changed the bandpass shape more rapidly, but in a manner that there are no systematic differences between the different LO phases. The outcome of this is shown in Fig.\,\ref{figfastvaryingbp} and Fig.\,\ref{figfastvaryingbpquality}.  
The LSFS method could not reconstruct the signal and gain curve. The residual is smooth but can only be described by a high-order polynomial. In fact, by computing the RMS with respect to a third-order polynomial we end up with significantly increased noise values and $\sigma$.
\begin{figure}[!t]
\plotone{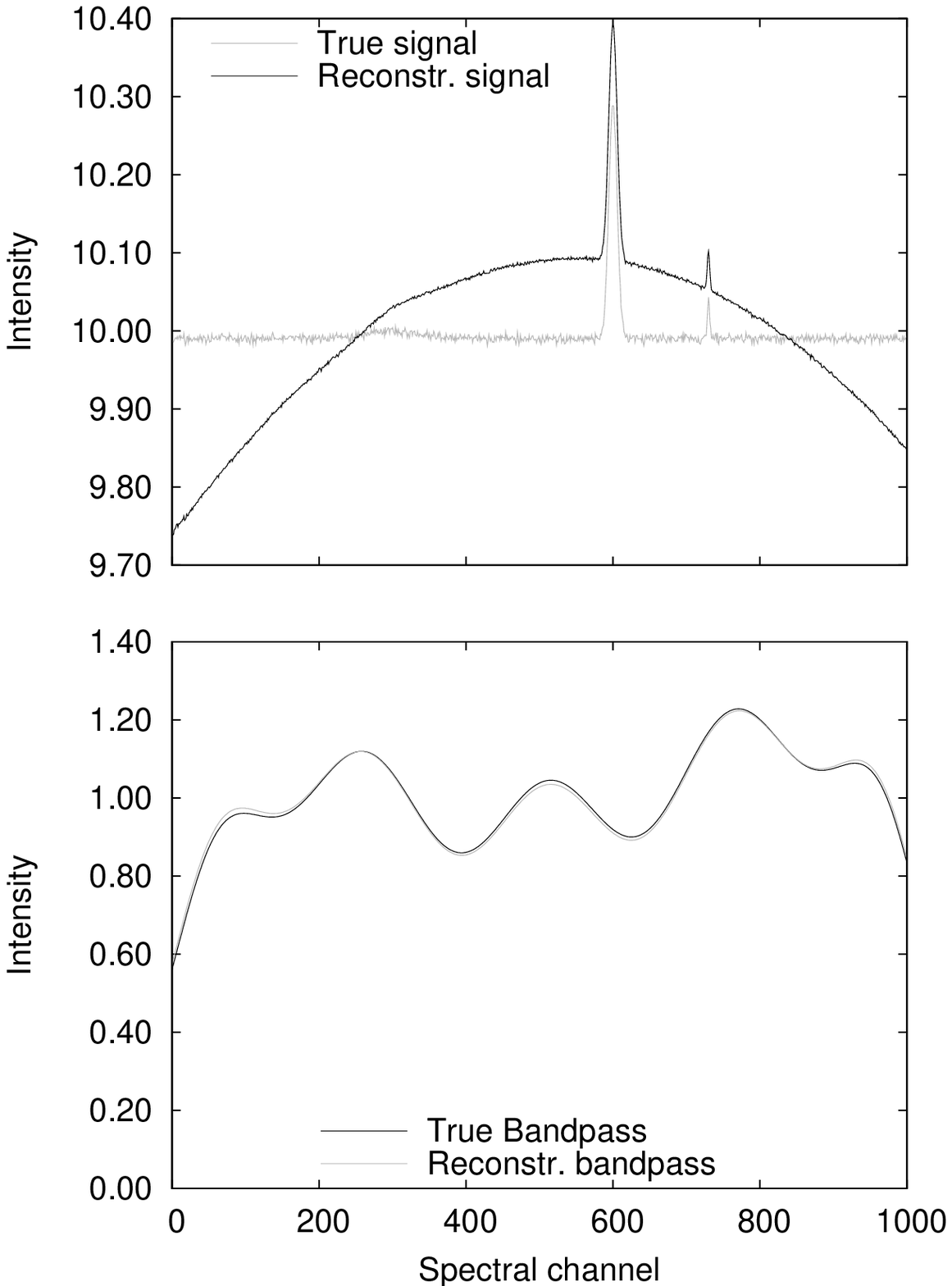}
\caption{In case of a systematic change of the bandpass shape which is due to the shift frequency (see text) LSFS fails to reconstruct the signal (top) and bandpass (bottom); see Fig.~\ref{figbpchangingduetofreqquality} for the quality indicators. }
\label{figbpchangingduetofreq}
\end{figure}

\begin{figure}[!t]
\plotone{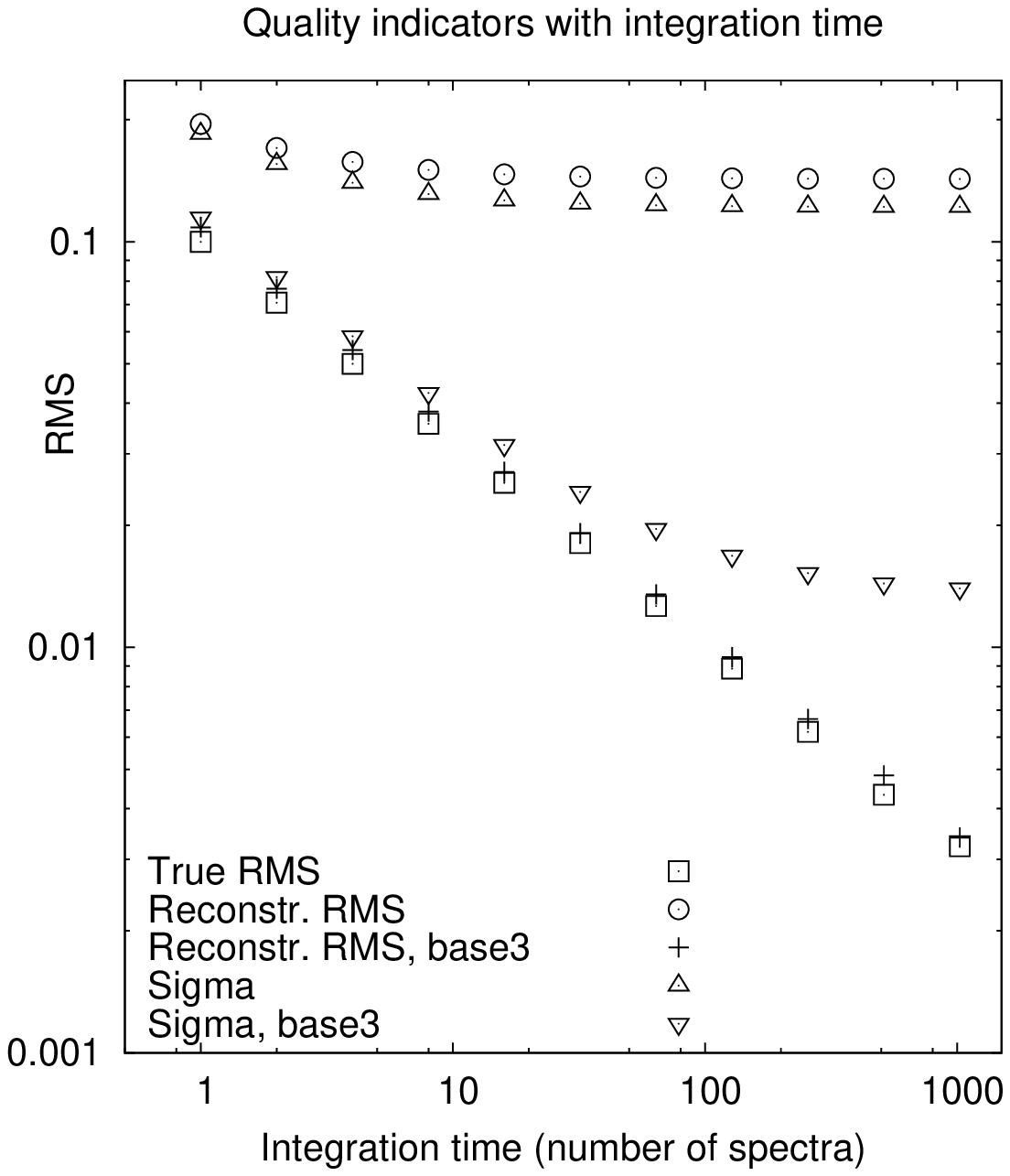}
\caption{Quality indicators for a systematic change of the bandpass shape which is due to the shift frequency (see text). The residual gain curve can --- to some extent --- be described by a low-order polynomial, but after integration of ~100 spectra the $\sigma$ value (see Fig.~\ref{figsensitivity} for the explanation of the symbols) no longer decreases.}
\label{figbpchangingduetofreqquality}
\end{figure}

For completeness, we also changed the shape of the bandpass in a systematic way by multiplying each bandpass with a linear function which slightly drops off towards higher frequency (negative slope). The slope of this function was steeper with higher shifting frequencies. This should mimic one of our early test observations with a digital fast-fourier-transform (DFFT) spectrometer prototype \citep{stanko05,winkel07}, where we were forced to use an LO frequency far off any specifications. We combined this systematic error with a slowly overall change of the bandpasses; see Fig.\,\ref{figbpchangingduetofreq} and Fig.\,\ref{figbpchangingduetofreqquality}. This time, the outcome was slightly better --- calculating RMS and $\sigma$ with respect to a 3rd-order polynomial leads to acceptable results in case of the signals RMS. However, the value of $\sigma$ does not decrease significantly after the summation of about 100 spectra.  At the end the noise is about a factor of four higher than expected. The RMS level of the reconstructed signal is not significant increased compared to the undisturbed case. Without subtracting a baseline the RMS and $\sigma$ values are even nearly independent on integration time.

We note that the latter two cases of very strong bandpass instabilities are far from any realistic scenario at modern radio telescopes. IF filter devices may have response to temperature and frequency variations but on a much smaller scale than we used to test the robustness of LSFS against those instabilities. In case of slowly varying gain curves the LSFS performs as good as without bandpass variations. 

\subsection{Continuum sources}

When mapping a region of the sky one often encounters the situation that continuum sources contribute significantly to the observed signal. Using the typical in-band frequency-switching algorithms, we have to assume that the spectra of these sources are sufficiently flat not showing any significant difference in the two switching phases. LSFS is much less dependent on this assumption as we switch only by a small fraction of the total bandwidth. But the greatest advantage of LSFS in this context is that the continuum signal will be part of the recovered signal spectrum, when using our normalization scheme. As spectrometer bandwidths have grown up to hundreds of MHz or even GHz nowadays it has become possible to also map continuum sources `for free' within a spectroscopic observation.
\begin{figure}[!t]
\plotone{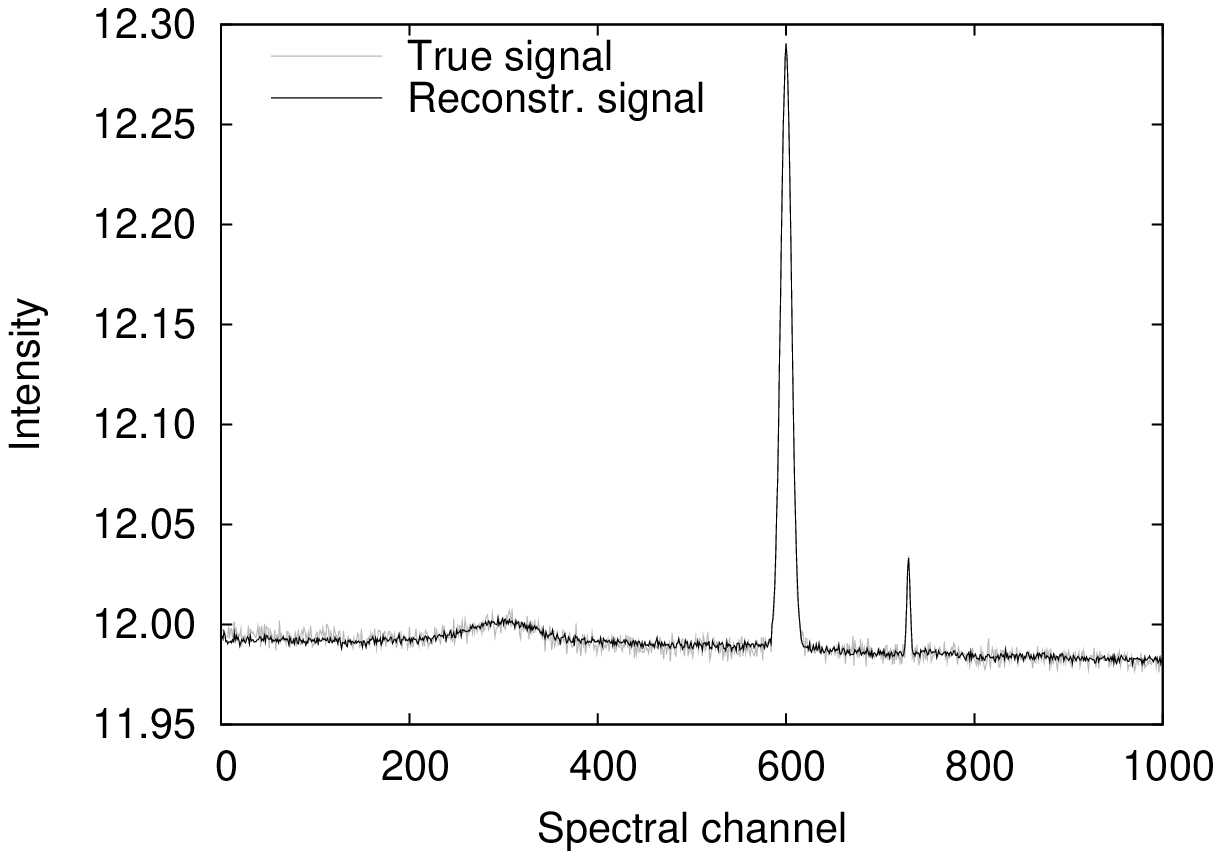}
\caption{The presence of a continuum source does not have negative influence on the outcome of the LSFS algorithm. The signal contains a continuum source of spectral index $\alpha=-2$ (top). Both spectral and continuum emission are well recovered as the quality indicators (see Fig.~\ref{figcontinuumsourcequality}) reveal. }
\label{figcontinuumsource}
\end{figure}

\begin{figure}[!t]
\plotone{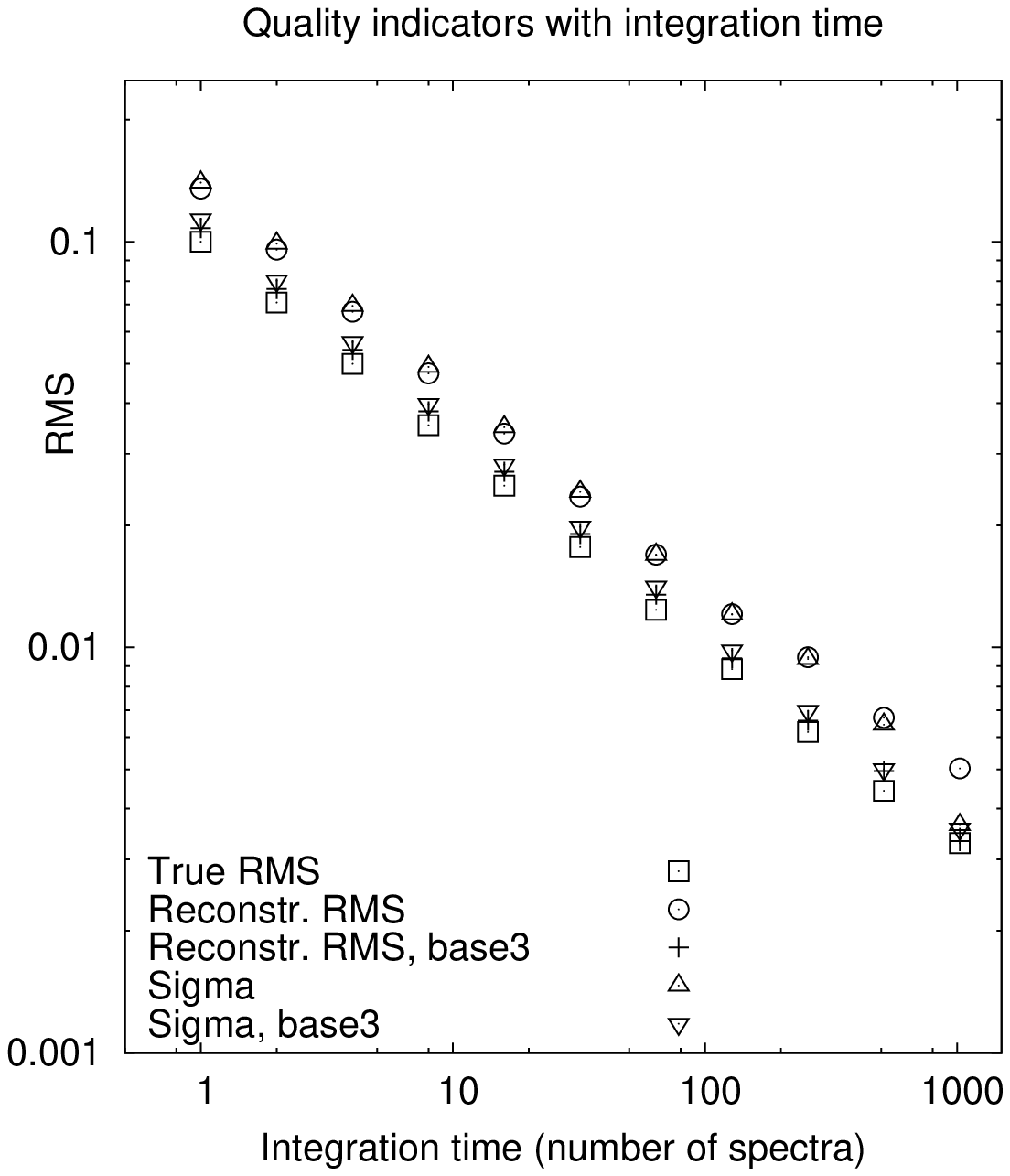}
\caption{Quality indicators (see Fig.~\ref{figsensitivity} for the explanation of the symbols) are not sensitive against continuum sources in the data. There is no significant increase of the RMS or $\sigma$ compared to the undisturbed case. }
\label{figcontinuumsourcequality}
\end{figure}

Fig.\,\ref{figcontinuumsource} shows the result for the case that a continuum source is superposed to the spectral lines. Its intensity is described by
\begin{equation}
I_\nu=A\left( \frac{\nu}{\nu_0} \right)^\alpha
\end{equation}
with spectral index $\alpha=-2$ and amplitude $A=2$ assuming $\nu_0=1420\,\mathrm{MHz}$ and a frequency resolution of 50\,kHz ($\delta v\approx10\,\mathrm{km\,s}^{-1}$) per spectral bin. Both the continuum signal as well as the spectral lines were nicely recovered, as the quality indicators (Fig.\,\ref{figcontinuumsourcequality}) show no increase in the RMS or $\sigma$ values.

\begin{figure}[!t]
\plotone{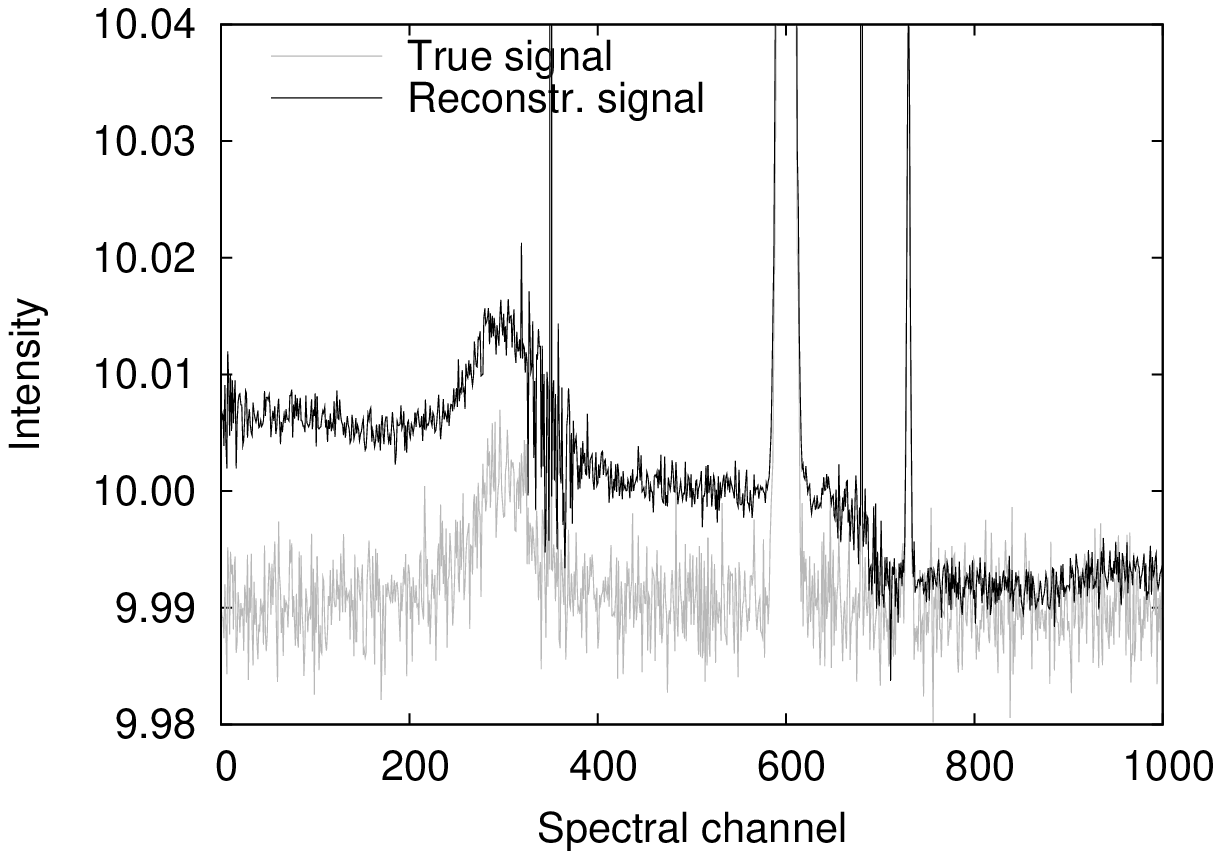}
\caption{RFI signals can have a severe effect on the solution of the LSFS. We added in spectral channels 350 and 680 a narrow-band interference signal. The fast-varying nature of the RFI signals added causes strong distortions of the reconstructed signal. For better visualization the plot shows a zoom-in. Fig.~\ref{figwithrfiquality} contains the quality indicators.}
\label{figwithrfi}
\end{figure}

\begin{figure}[!t]
\plotone{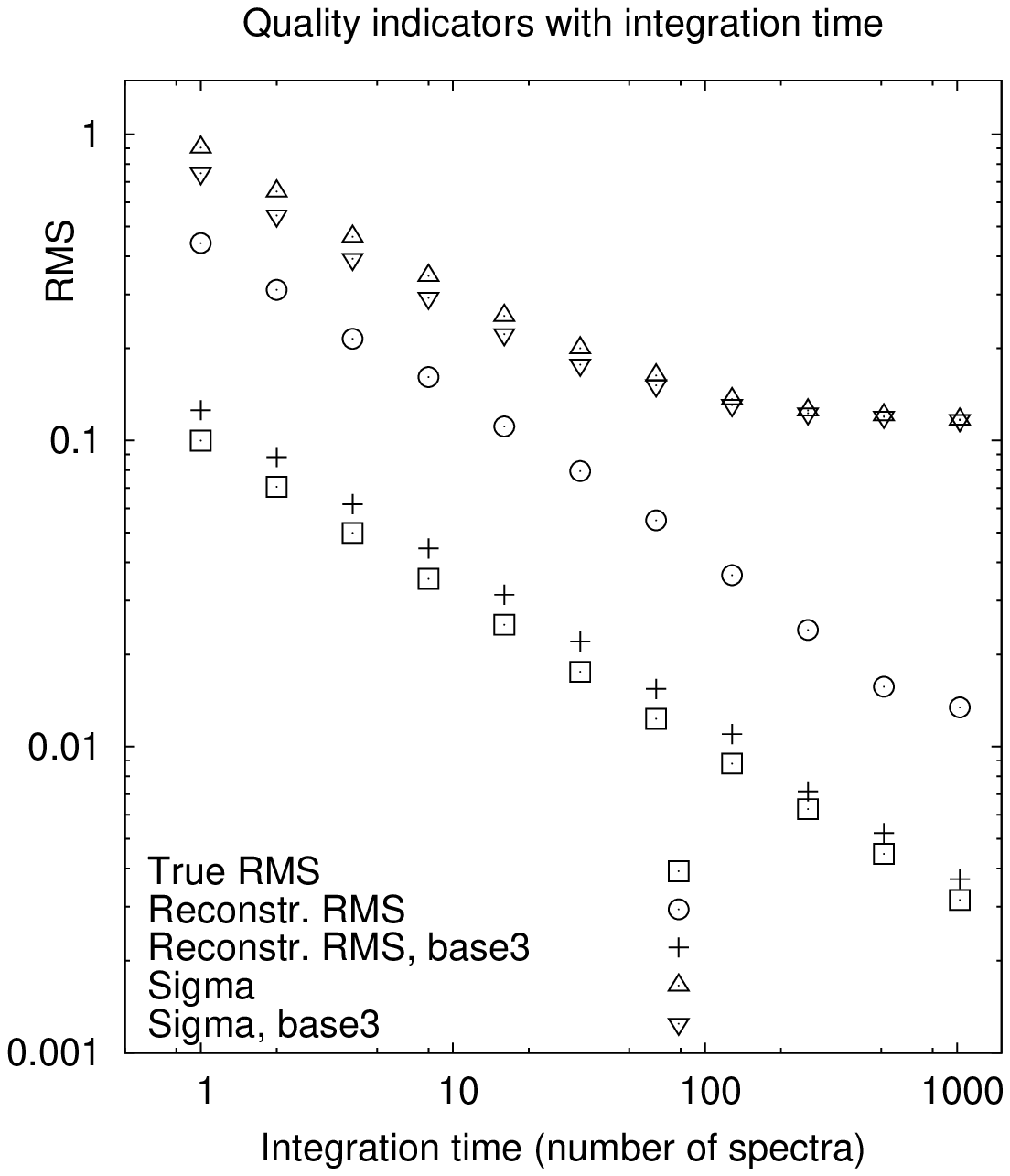}
\caption{The quality indicators (see Fig.~\ref{figsensitivity} for the explanation of the symbols) are very sensitive to RFI signals. After subtracting a third-order baseline the signals noise level is increased by about 20\%, while $\sigma$ is even increased by a factor of $\gtrsim4$.}
\label{figwithrfiquality}
\end{figure}

\subsection{Radio frequency interference}

One of the key properties of each data reduction pipeline used in radio astronomy today is the capability to deal with radio frequency interferences (RFI). These artificial signals are in general variable on timescales down to $\mu$s. Therefore, one of the most interesting analyses in this section is the impact of such interferences on the LSFS method. For simplicity, we started by adding two narrow-band interferences whose amplitudes obey a power law. This is --- at least at the 100-m telescope at Effelsberg --- one of the most common types of interference.

As the LSFS algorithm assumes the signal to be stable, it was not surprising that the result is practically useless. Due to the coupling of channels with different frequency shifting, we end up with a number of contaminated spectral channels which is higher than the initial number of channels affected; see Fig.\,\ref{figwithrfi}. The only solution is actually to address the RFI problem before performing the LSFS. 

\cite{winkel07} presented an algorithm which detects interferences down to the $\lesssim 4\sigma_{\textrm{rms}}$ level. Having detected interference peaks, one can flag these data in order to exclude them from the computation. Flagging data points is equivalent to projecting the correlation matrix in Eq.\,(\ref{eqlinearmatrix}) to a subspace which does not contain contaminated spectral channels. This, however, would require to recompute the SVD of the matrix each time the spectral channels containing RFI would change. This is far from practical as computing the SVD for 1024 spectral channels and 8 LO frequencies takes at least a few minutes on a modern PC. 
\begin{figure}[!t]
\plotone{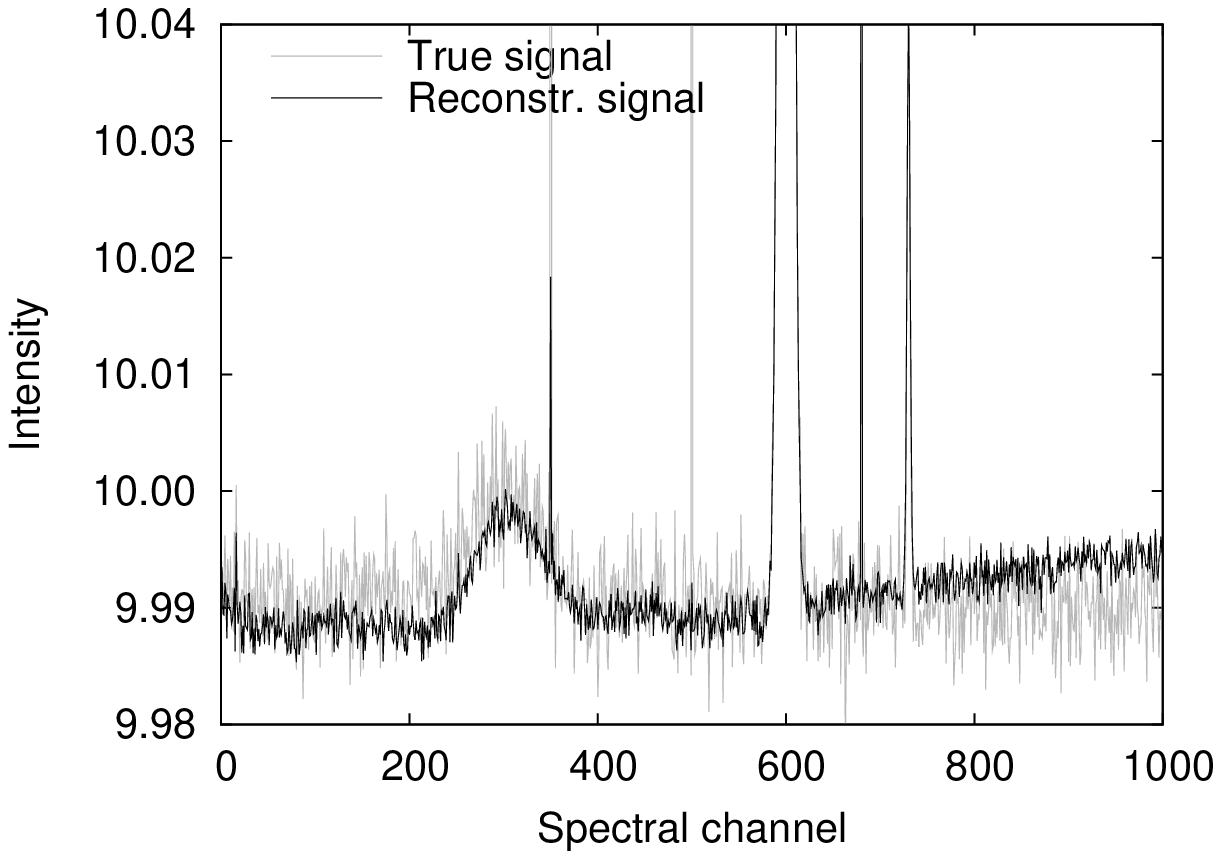}
\caption{Proper handling of RFI contaminated data points allows reconstruction of the signal. In the signal domain residual RFI peaks remain, but have less amplitude and no measurable influence on their environment as this was seen in Fig.\,\ref{figwithrfi}. We added three narrow-band interferences. The RFI signal in spectral channel 350 was added in all LO phases except for LO 1 and 2. The signal in spectral channel 500 affected every second LO phase while the signal in channel 680 was added in all phases. It turns out, that if an RFI signal is not persistent for a whole LO cycle the unaffected data points in the associated spectral bin can even be sufficient to reconstruct the signal without artifacts. The less LO phases are affected, the less impact of the RFI on the reconstructed signal is visible.  }
\label{figwithrfiflagging}
\end{figure}

\begin{figure}[!t]
\plotone{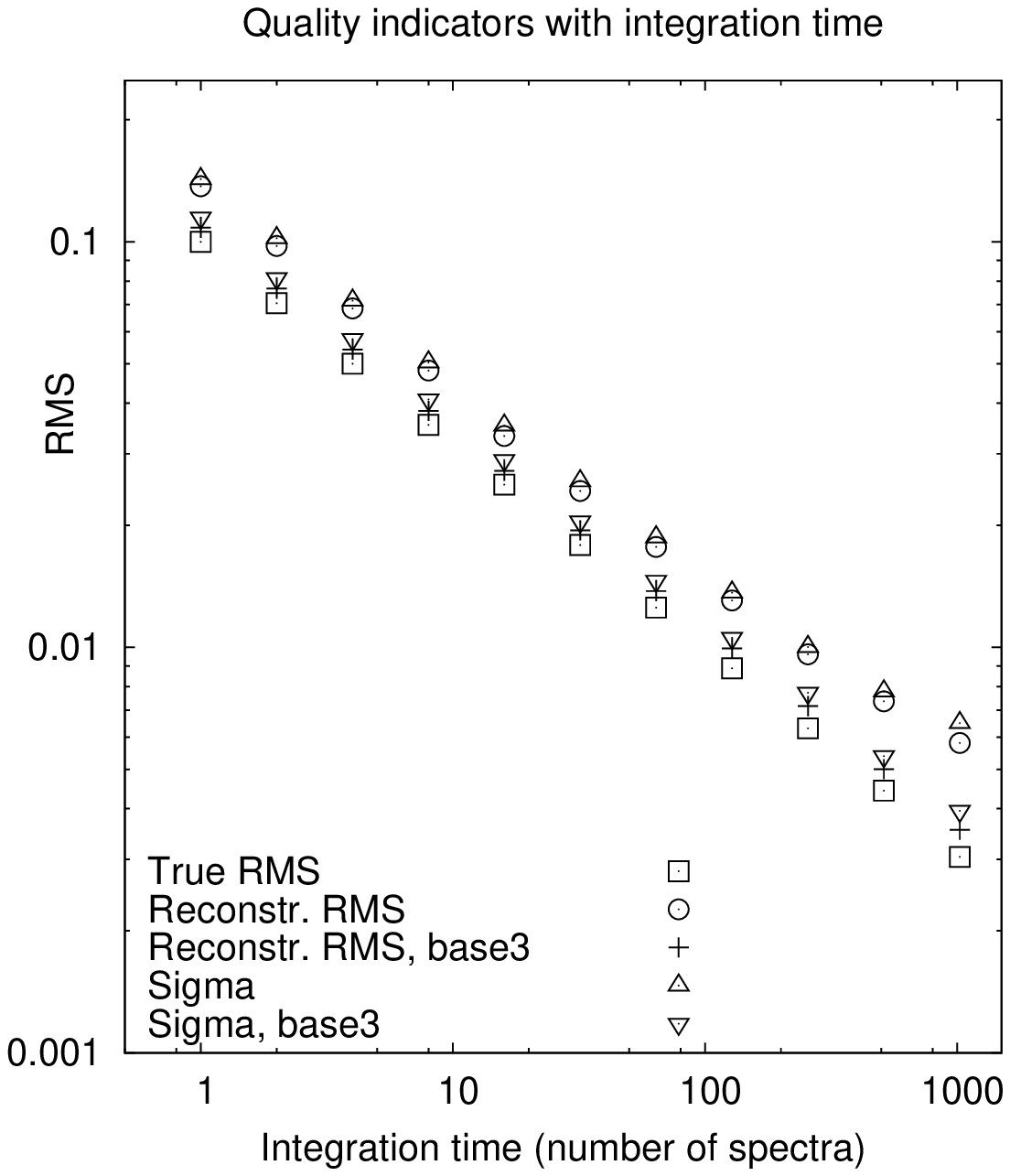}
\caption{Using the flagging scheme to suppress distortions by RFI signals provides noise level (RMS) values (bottom; see Fig.~\ref{figsensitivity} for the explanation of the symbols), which are only $\sim14\%$ higher and a value of $\sigma$, which is  $\sim25\%$ higher than theoretically. }
\label{figwithrfiflaggingquality}
\end{figure}

By far easier is the following alternative: setting all spectral channels containing an RFI signal (those are of course different channel numbers for different shifting frequencies) in $\mbox{\boldmath $p$}$ to zero. 
Of importance is here a robust calculation of the mean signal strength by dropping all disturbed spectral channels. Otherwise, the gain factors would depend on the actual strength of the RFI signals. 

As our algorithm is not able to find RFI signals hidden in the noise (though an iterative scheme may be possible, were one performs the search for interferences at different integration levels), we only set spectral channels to zero which contain an interference signal of $\geq 4\sigma_{\textrm{rms}}$. We added three narrow-band RFI signals whose amplitudes obey a power law with spectral index $\nu=-1.5$. The leftmost signal at spectral channel~250 was persistent in all LOs except 1~and~2. The signal at channel~600 was only added in every second LO and the rightmost interference at channel~680 was added for each LO. 

The outcome is shown in Fig.\,\ref{figwithrfiflagging} and Fig.\,\ref{figwithrfiflaggingquality}. The bandpass was well recovered. However, each RFI leaves behind some `fingerprint' in the reconstructed signal, the residual strength of which obviously depends on the number of affected LOs. These ``left-overs'' are nevertheless easy to handle as the spectral channels and LOs containing RFI are more or less known (otherwise the flagging would not have been possible).

\begin{figure}[!t]
\plotone{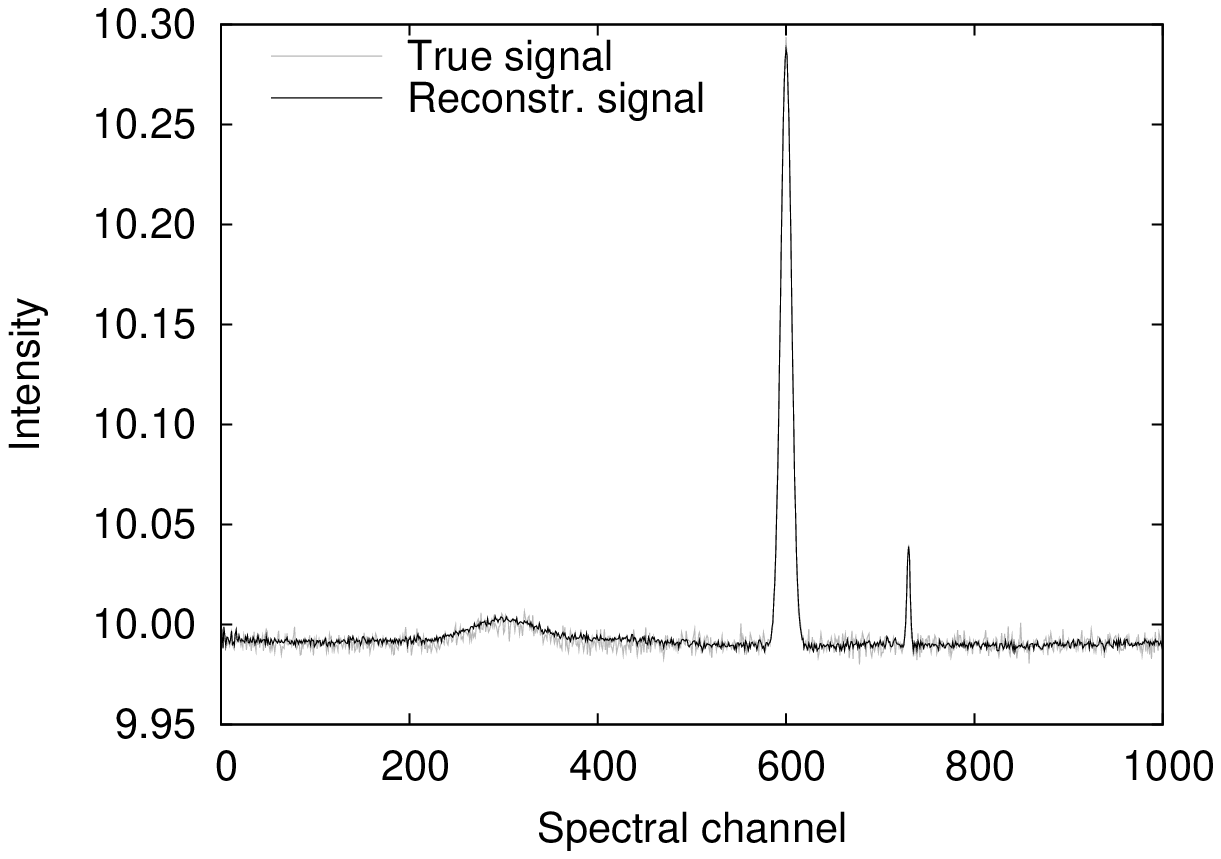}
\caption{LSFS for a broadband interference signal. As only one LO frequency is affected, the signal and bandpass could be well recovered. Fig.~\ref{figwithrfiflaggingbroadbandquality} shows the quality indicators which are only slightly increased compared to the undisturbed case.}
\label{figwithrfiflaggingbroadband}
\end{figure}

\begin{figure}[!t]
\plotone{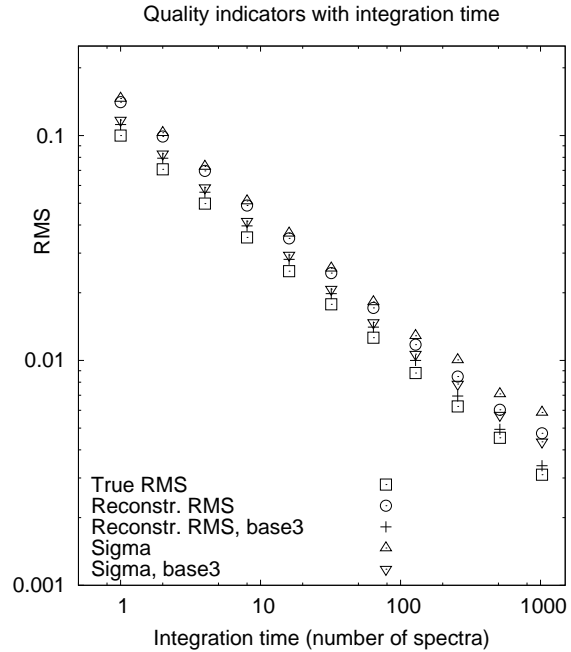}
\caption{Quality indicators (see Fig.~\ref{figsensitivity} for the explanation of the symbols) for spectra containing a broadband interference signal.}
\label{figwithrfiflaggingbroadbandquality}
\end{figure}

Implementing RFI flagging enables the analysis of the response of the LSFS to different types of RFI. During our measurements we rarely encountered broad-band events, which last for only a second or less but affect several hundred spectral channels \citep{winkel07}.  Fig.\,\ref{figwithrfiflaggingbroadband} and Fig.\,\ref{figwithrfiflaggingbroadbandquality} show the result for affecting the 4th LO within spectral channels 200 to 400 --- the reconstruction was successful when using our flagging scheme --- it was not otherwise (not shown here). The intensities of the broad-band signal are drawn from a power law but lie within $4\ldots20\sigma_{\textrm{rms}}$. We added the interference onto each spectrum of the 4th LO which would hardly be the case for a real observation (this type of RFI is rare). 

\section{Computational efficiency}\label{seccomputingeff}
When we started the analysis by implementing the LSFS within the C programming language we chose for the sake of simplicity the SVD algorithms delivered with the GNU Scientific Library (GSL)\footnote{\url{http://www.gnu.org/software/gsl/}}. They make use of the modified Golub-Reinsch algorithm. But in our case we can save a lot of computing time by using the fact that we have a sparse matrix. There exist a few libraries (mainly for FORTRAN) which use the Lanczos (SVD) algorithm for sparse matrices. We used the \texttt{las2} routine from SVDPACKC\footnote{\url{http://www.netlib.org/svdpack/}} through the interface library SVDLIBC\footnote{\url{http://tedlab.mit.edu/ \textasciitilde   dr/SVDLIBC/}}. Table\,\ref{tabcomputingtime} lists run-times of the pure SVD computation for different $I$, using both methods. The \texttt{las2} algorithm is about an order of magnitude faster, which means it can calculate the SVD of a two times larger matrix within the same time (as the SVD computation scales as $I^3$). Therefore, it is the preferred method for large values of $N\cdot I$. Note also, that the main memory needed scales roughly as $NI^2$. For the largest of our problems ($I=2048,\,N=8$) a 1-GB-machine was barely sufficient using double precision arithmetic. 

As the SVD needs only to computed once per LO setup the more important contribution to computing times needed is due to the LSFS calculation itself. Based on our experience in many cases the convergence is reached after few ($\lesssim5$) steps. When confronted with RFI etc. this increases up to 20 or more iterations until convergence. To account for this we monitor changes of the solution signal and break the iteration after the solution has stabilized. Based on different numbers of steps needed, the computation of the LSFS ($N=8$, $I=1024$) takes $\sim0.05\,\mathrm{s}$ per iteration step on a modern desktop PC (2.0 GHz, x86). We already used an optimized BLAS library which makes use of SSE or equivalent features of modern x86 cpu's meaning that there is probably not much potential to speed up the computation of the LSFS.

We also played around a little bit with the compiler extension OpenMP\footnote{\url{http://www.openmp.org/}} to parallelize the LSFS for use on multi-processor/core machines. This could improve the run-time by about 25\% on a Dual-Xeon machine and about 15\% on a Dual-Core processor. The maximum speed-up one could expect would be a factor of two. In fact, the LSFS computation depends mainly on the multiplication of the (huge) correlation matrix with the input vector. Here the memory bandwidth has large impact on the overall speed which is possible. 
\begin{table}
\caption{Computing times needed to calculate the SVD using different algorithms. }
\label{tabcomputingtime}
\centering
\begin{tabular}{lllllll}\tableline\tableline
$I$&$N$&rows&cols&matrix&\multicolumn{2}{c}{time (s)\tablenotemark{a}}\\ 
&&&&density&sparse&gsl\\\tableline
128&	8&	1025&	300&	0.72&	$\simeq1$&	2\\ 
256&	8&	2049&	556&	0.39&	3&	20\\ 
512&	8&	4097&	1068&	0.20&	28&	168\\ 
1024&	8&	8193&	2092&	0.10&	220&	2454\\ 
2048&	8&	16385&	4140&	0.05&	2573&	N/A\\ 		
\tablenotetext{a}{Using a 2.0 GHz x86 CPU.}
\end{tabular}
\end{table}

\section{Summary}\label{secsummary}
In this paper we analyzed the statistical behavior of the LSFS method as a function of integration time as well as the robustness of this new method against various potential sources of errors as RFI signals and gain curve instabilities. It turned out that LSFS will provide very good solutions in most cases. However, in case of RFI the solution is strongly disturbed, rendering LSFS useless. We developed a flagging scheme which is able to deal with interferences if there is a detection database containing accurate information where (in time and frequency) RFI signals were present.

Mild bandpass instabilities are no problem at all but very fast variations can cause moderate to severe distortions of the reconstructed gain curve. The latter, however, are far from realistic scenarios at modern radio telescopes, making LSFS the best choice even when confronted with ugly (but nearly time-independent) bandpass shapes. A strong advantage of LSFS versus common frequency switching methods is that there is only a small frequency shift needed which results in much less bandpass variations at all. 

We also have shown that the LSFS will fail in presence of very strong emission lines as would be the case for example in galactic \ion{H}{1} research due to the strong Milky Way \ion{H}{1} line emission of the disk. Here the linear order approximation is broken. We presented a possible workaround by remapping the signal. 

While we could not work out a significant speed increase for the pure LSFS computation, we at least propose the usage of a specialized algorithm to compute the SVD of the correlation matrix. Such an algorithm turns out to operate an order of magnitude faster for sparse matrices, which in turn allows the computation of such a matrix for twice the number of spectral channels within the same computing time.

\acknowledgments

\textit{Acknowledgments.} We would like to thank Carl Heiles for making his manuscript about the LSFS method available to us. Thanks also to Tobias Westmeier for many useful comments. Benjamin Winkel was supported for this research through a stipend from the International Max Planck Research School (IMPRS) for Radio and Infrared Astronomy at the Universities of Bonn and Cologne.






\bibliographystyle{aa}
\bibliography{references}

\end{document}